\documentclass[11pt,a4paper]{article}
\pdfoutput=1
\usepackage[a4paper,margin=1in]{geometry}
\usepackage{setspace}
\usepackage{authblk}
\usepackage{subcaption}
\usepackage{float}
\usepackage{amsmath,amssymb}
\usepackage{graphicx}
\usepackage[backend=biber,style=numeric,sorting=none]{biblatex}
\usepackage{hyperref}

\begin{document}

\title{Nitrogen-induced ELM suppression and confinement improvement in the EAST tokamak with a full metal wall}

\author[1]{Jingyan Hu}
\author[2,\thanks{Email: pshi@ipp.ac.cn}]{Peng Shi}
\author[3]{Chu Zhou}
\author[2]{Jinyue Liu}
\author[2]{Gongshun Li}
\author[2]{Kangning Geng}
\author[4]{Yiren Zhu}
\author[2]{Kedong Li}
\author[2]{Hailin Zhao}
\author[2]{Xiang Jian}
\author[3]{Ge Zhuang}

\affil[1]{School of the Gifted Young, University of Science and Technology of China, Hefei 230026, China}
\affil[2]{Institute of Plasma Physics, Chinese Academy of Sciences, Hefei 230031, China}
\affil[3]{School of Nuclear Science and Technology, University of Science and Technology of China, Hefei 230026, China}
\affil[4]{Southwestern Institute of Physics, PO Box 432, Chengdu 610041, China}

\maketitle

\begin{abstract}
This paper reports the achievement of an ELM-free H-mode regime with confinement improvement enabled by nitrogen (N$_2$) seeding on the Experimental Advanced Superconducting Tokamak (EAST) with a full metal wall. Following N$_2$ injection, large Edge-Localized Mode (ELM) bursts are completely suppressed, while global energy confinement is significantly enhanced, with the $H_{98(y,2)}$ factor increasing from approximately 0.9 to 1.2. A distinct edge coherent mode (ECM), localized at the pedestal foot $(\psi_N\sim0.99)$, is identified using O-mode Poloidal Correlation Reflectometry and AXUV diagnostics. This mode operates within a frequency range of 20–50~kHz with a poloidal wavenumber of $k_\theta \approx 0.54~$cm$^{-1}$. Linear gyrokinetic simulations performed with the CGYRO code reveal a dominant instability that quantitatively matches the experimental measurements. Detailed scans of parameters identify this mode as a Dissipative Trapped Electron Mode (DTEM). The energy and particle transport driven by this pedestal-foot DTEM effectively regulates the edge gradients, preventing the pedestal from crossing the Peeling-Ballooning stability boundary and sustaining a stationary ELM-free state. These findings provide a physical basis for an integrated scenario to maintain high confinement and protect plasma-facing components in future steady-state fusion reactors.
\end{abstract}

\section{Introduction}
The High-Confinement Mode (H-mode), characterized by elevated core density and temperature~\cite{wagner1982regime}, represents a primary operational scenario for magnetic confinement fusion devices aiming to achieve the requisite fusion triple product $n T \tau_E$ and ignition condition~\cite{lawson1957some,connor1998magnetohydrodynamic,mcnally1973nuclear}.

In H-mode, an edge transport barrier known as the pedestal forms~\cite{wagner1982regime,snyder2004elms}, which significantly reduces turbulent transport and sustains the high core confinement~\cite{urano2014pedestal}.

However, the steep pressure gradients and associated bootstrap currents in the pedestal can trigger Peeling-Ballooning (PB) instabilities~\cite{connor1998magnetohydrodynamic,urano2014pedestal}.

As the pedestal evolves, pressure and current density profiles eventually breach the P-B stability boundary and trigger Edge-Localized Modes (ELMs)~\cite{snyder2004elms}.

ELMs expel bursts of particles and energy into the Scrape-Off Layer (SOL) and induce transient heat fluxes onto plasma-facing components (PFCs)~\cite{huijsmans2015modelling}. In reactor-scale plasmas, these transients can impose damaging heat loads on divertor targets and other PFCs, posing a significant threat to the longevity of steady-state reactors~\cite{leonard1999impact,herrmann2003stationary}. The problem is expected to become more restrictive in future burning plasma devices (e.g., ITER)~\cite{snyder2009pedestal}, where smaller normalized ion gyro-radii ($\rho_i$) and collisionality can reduce diamagnetic stabilization of medium$n$ P-B modes~\cite{snyder2004elms,huijsmans2015modelling}. Therefore, developing integrated control scenarios that suppress ELMs while maintaining high confinement remains a primary goal for steady-state fusion operation.

Currently, several active ELM-control methods have been developed, including pellet pacing~\cite{pacher2011elm}, Resonant Magnetic Perturbations (RMP)~\cite{fitzpatrick2020modeling}, and impurity powder injection (e.g., Lithium, Boron)~\cite{sun2021suppression}. While effective in certain regimes, these methods face challenges regarding engineering complexity, operational windows (e.g., $q_{95}$ restrictions for RMP), and compatibility with reactor-grade fueling or wall-conditioning requirements~\cite{lang2013elm,huijsmans2015modelling}.

Impurity gas seeding offers a flexible alternative. In future metallic-wall reactors like ITER, low-Z species such as Nitrogen (N$_2$), Neon (Ne), and Argon (Ar) are generally categorized as low-$Z$ impurities, distinct from high-$Z$ intrinsic metallic impurities. Seeding these low-$Z$ impurities can increase edge radiation and help manage steady-state power exhaust. In current tokamaks, low-Z impurity injection has been shown to successfully enable access to ELM-mitigated or ELM-suppressed regimes, and sometimes can achieve divertor detachment simultaneously~\cite{li2023compatibility,rapp2002elm,casali2025achievement,lin2022physical,zhong2019impact}. However, its impact on pedestal stability and plasma confinement is not universal~\cite{li2023compatibility,lin2022physical,casali2025achievement}. Optimized seeding can can access small- or no-ELM regimes and even improve confinement~\cite{li2023compatibility}, whereas excessive radiation, fuel dilution, or unfavorable edge profile changes can degrade performance~\cite{rapp2002elm,lin2022physical,casali2025achievement}.

This non-universality points to an unresolved physics question: how impurity seeding modifies the pedestal transport channels that determine whether the plasma relaxes gently or enters large ELM cycles. Previous EAST studies have shown that specific turbulence, such as edge coherent modes~\cite{wang2014new} and high-frequency broadband turbulence~\cite{xu2026turbulence}, can provide continuous transport and regulate ELM activity in long-pulse H-mode plasmas. 

Here we report on a nitrogen-induced ELM-free H-mode regime on EAST with a full metal wall. After Nitrogen injection, large ELMs are suppressed while the global confinement improves, with $H_{98(y,2)}$ increasing from about 0.9 to 1.2. Using O-mode poloidal correlation reflectometry (PCR), absolute extreme-ultraviolet (AXUV) arrays and magnetic diagnostics, we identify a predominantly electrostatic edge coherent mode. Distinct from the typical Edge Harmonic Oscillations (EHO)~\cite{burrell2001quiescent} or Edge Coherent Modes (ECM)~\cite{guo2014recent} often observed in the steep-gradient region, the ECM found here is localized at the pedestal foot. This mode operates within a frequency range of 20–50~kHz, with a poloidal wavelength of approximately 11.7~cm. 

Linear gyrokinetic simulations with CGYRO show that the measured wavenumber, propagation direction and eigenfunction parity are consistent with a dissipative trapped electron mode (DTEM). Parameter scans further show that this mode is driven by the electron temperature gradients and is destabilized by the enhanced collisionality associated with nitrogen seeding. However, the ELM-free phase is accompanied by a stronger temperature pedestal and improved global confinement. ELITE analysis also indicates a higher linear peeling-ballooning growth rate after ELM suppression. These observations rule out a simple pressure-gradient-clamping explanation. We therefore propose that the nitrogen-induced ECM suppresses ELMs through nonlinear interaction with peeling-ballooning dynamics, consistent with coherent-mode-mediated ELM mitigation mechanisms reported in recent EAST simulations~\cite{li2022simulation} and multiscale tokamak-edge studies~\cite{li2025multi}.

\section{Experimental Setup}

The experiments were conducted on the Experimental Advanced Superconducting Tokamak (EAST)~\cite{li2021experimental}, which was operated in a single-null divertor configuration. This device is equipped with ITER-like water-cooled tungsten (W) monoblocks as plasma-facing components (PFCs) for handling high heat fluxes. To access the H-mode and sustain the plasma performance in this discharge, auxiliary heating was provided by a combination of Electron Cyclotron Resonance Heating (ECRH) and Neutral Beam Injection (NBI).

\begin{figure}[H]
    \centering
    \includegraphics[width=0.65\linewidth]{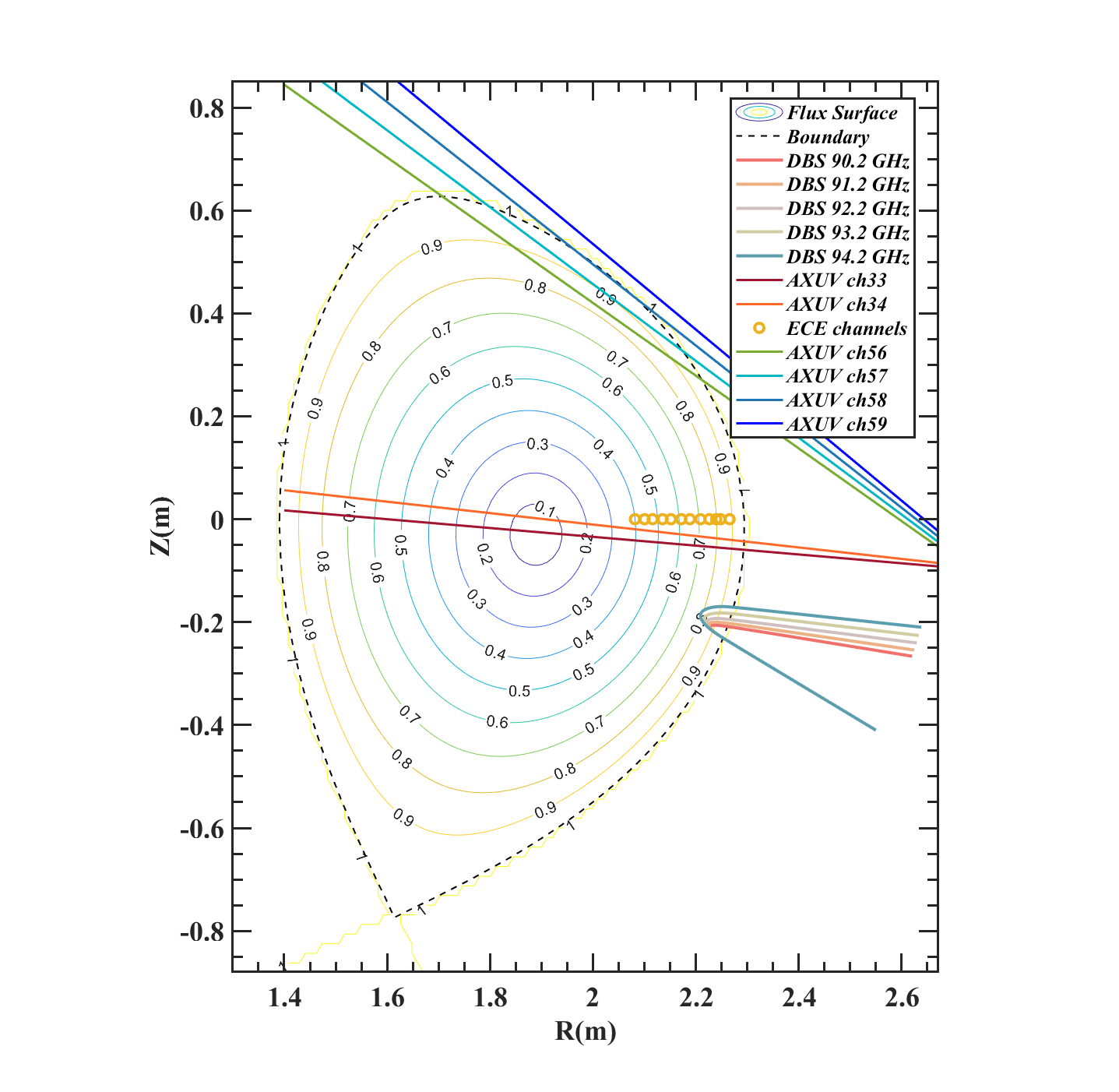}
    \caption{Overview of the experimental diagnostics and equilibrium configuration. The reflectometry for density measurements is located 3 cm above the midplane. ECE channels are positioned near the low-field side (LFS). The flux surfaces represent the reconstructed equilibrium for discharge \#153901.}
    \label{fig:diagnostics}
\end{figure}

To comprehensively evaluate the plasma response to nitrogen seeding, a suite of high-resolution diagnostic systems was employed, as illustrated in Fig.~\ref{fig:diagnostics}. Electron density ($n_e$) profiles were provided by the Density Profile Reflectometry (DPR) system~\cite{wang2019development}. Electron temperature ($T_e$) profiles were reconstructed from Thomson scattering measurements~\cite{zang2011upgraded} and the Electron Cyclotron Emission (ECE) data~\cite{liu2018overview}.

Pedestal fluctuations and coherent modes were diagnosed with complementary microwave systems covering different radial locations. Doppler Backscattering (DBS) system, operating in X-mode over the W-band (75–110~GHz)~\cite{feng2019five}, probed the pedestal top and steep-gradient region. Poloidal Correlation Reflectometry (PCR) system, operated in O-mode over the K-Ka band (20.4–40~GHz)~\cite{xiang2018development} and the U-band (42.4–57.2~GHz)~\cite{zhou2021experimental,li2025experimental}, monitored the middle-to-lower pedestal with partial overlap with the DBS measurement region.

The radiative cooling and impurity penetration were assessed using absolute extreme ultraviolet (AXUV) photodiode arrays ~\cite{duan2012resistive} and extreme ultraviolet (EUV) spectroscopy~\cite{hou2024development}. The AXUV system comprises four horizontal arrays with a total of 64 channels, providing extensive coverage of the plasma cross-section. Furthermore, the N$_{VII}$ line emission intensity measured by EUV spectroscopy was used as a proxy for nitrogen penetration and ionization state in the bulk plasma. Additionally, ELM activity and magnetic fluctuations were monitored using $D_\alpha$ emission arrays and magnetic Mirnov coils.

\section{Phenomenology of ELM suppression and data analysis}

\subsection{Macroscopic evolution and pedestal profiles}

\begin{figure}[H]
    \centering
    \includegraphics[width=0.63\linewidth]{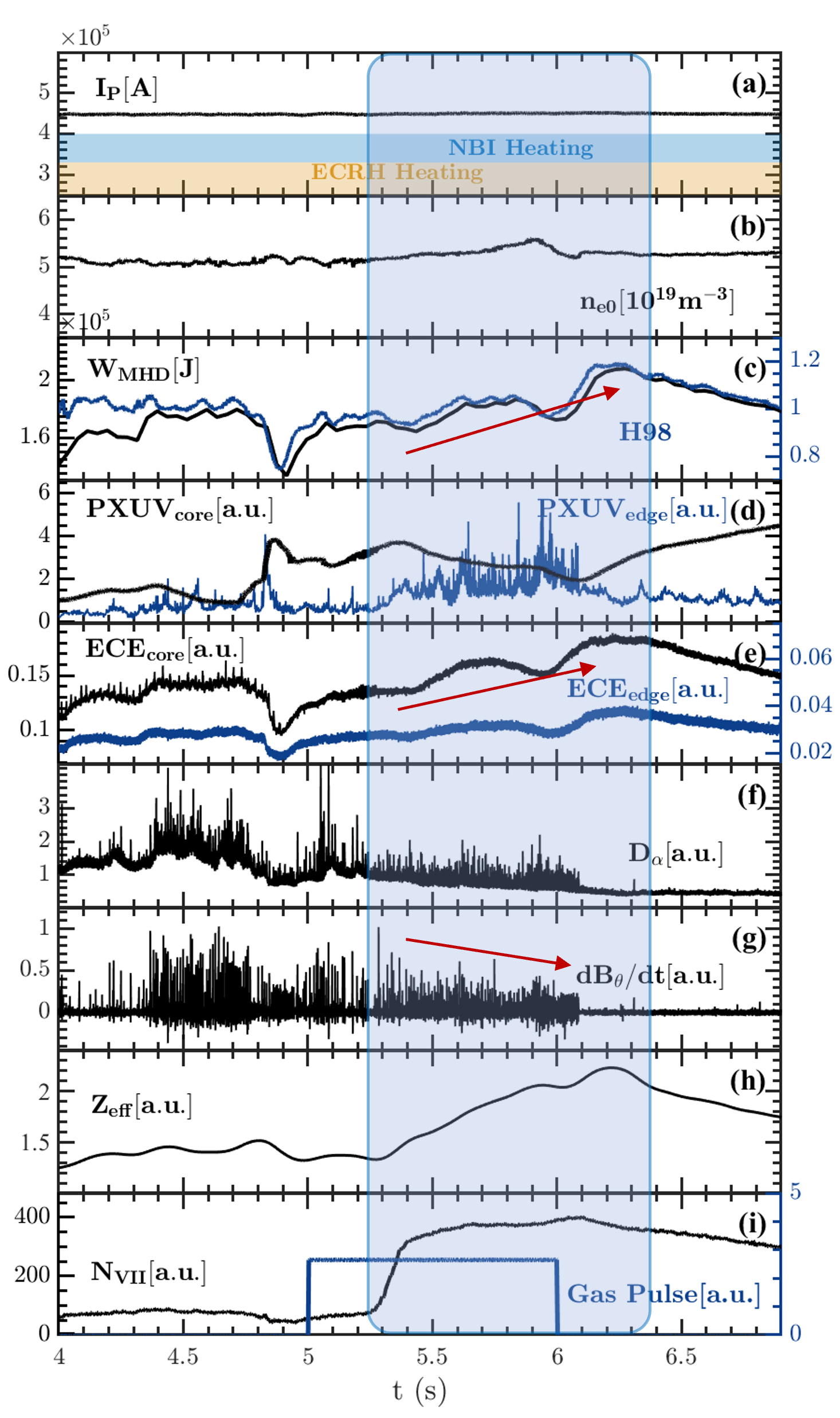}
    \caption{Time evolution of main parameters for EAST discharge \#153901: (a) Plasma current $I_p$ and auxiliary heating (NBI and ECRH), (b) line-averaged electron density $\bar{n}_e$, (c) plasma stored energy $W_{MHD}$ and confinement factor $H_{98(y,2)}$, (d) core and edge PXUV radiation, (e) core and edge ECE temperature signals, (f) divertor $D_\alpha$ emission, (g) magnetic Mirnov coil signal, (h) effective nuclear charge $Z_{eff}$, (i) N$_{VII}$ emission and N$_2$ gas puffing pulse.}
    \label{fig:discharge}
\end{figure}

The temporal evolution of key plasma parameters for the representative Nitrogen-seeded discharge (\#153901) is displayed in Fig.~\ref{fig:discharge}. The target H-mode plasma was operated with a plasma current of $I_p \approx 450~kA$, an edge safety factor of $q_{95} \approx 5.3$, a normalized energy confinement factor $H_{98(y,2)} \approx 1.0$ and a line-averaged electron density of $n_e \approx 5.2 \times 10^{19}$~m$^{-3}$ (corresponds to a Greenwald density fraction of $f_{GW} \approx 0.74$). Prior to impurity injection, the discharge exhibited ELMy H-mode characteristics.

To achieve and sustain the H-mode regime at high plasma beta, a total of 2.7~MW of auxiliary heating was applied, consisting of 1.35~MW of NBI and 1.35~MW of ECRH. The NBI system was active from $t = 4.0$~s to $10.0$~s, while the ECRH system provided continuous heating from $t = 0.0$~s to $10.9$~s, fully covering the impurity injection window.

Nitrogen seeding was applied using a localized impurity gas injection system. A gas mixture of 50\% N$_2$ and 50\% D$_2$ was injected from the low-field side midplane during the commanded puff interval of $t = 5.0$-$6.0$~s. The actual impurity arrival in the plasma was identified from the boundary AXUV channel 59 and the EUV nitrogen radiation signal. The increase of boundary AXUV and nitrogen radiation indicate that N$_2$ entered the plasma at approximately 5.25~s, corresponding to a delay of about 0.25~s from the gas inlet to the plasma. Therefore, the shaded region in Fig.~\ref{fig:discharge} denotes the time interval after impurity penetration into the plasma.

After nitrogen entered the plasma, the line-averaged density remained nearly unchanged under feedback control, whereas the core and edge radiation increased. ECE measurements show that both core and edge temperatures rose during the seeding phase. Consistently, the plasma stored energy $W_{MHD}$ increased from approximately 160~kJ to 200~kJ, and $H_{98(y2)}$ increased from 0.9 to 1.2. At the same time, the $D_\alpha$ spikes and Mirnov bursts were progressively weakened and then fully suppressed at approximately 6.1~s; this ELM-free phase persisted until about 6.95~s.

\begin{figure}[H]
    \centering
    \includegraphics[width=0.9\linewidth]{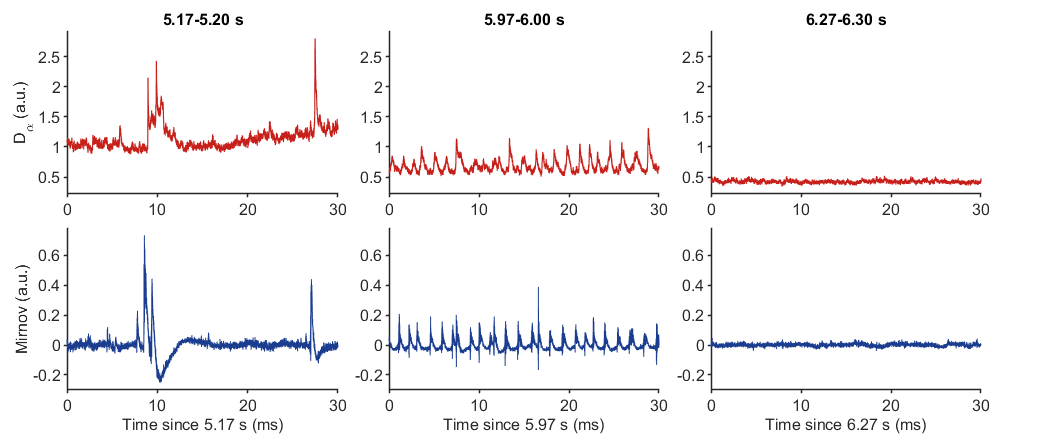}
    \caption{Expanded $D_\alpha$ and Mirnov signals in three representative time windows. The ELM activity changes from large-amplitude, low-frequency Type-I ELMs before impurity penetration (5.17-5.20~s) to small-amplitude, high-frequency Type-III ELMs around 6.0~s (5.97-6.00~s), and then to an ELM-free phase at 6.27-6.30~s.}
    \label{fig:represented}
\end{figure}

Fig.~\ref{fig:represented} further resolves the change in ELM behavior during the nitrogen-seeding sequence. Before impurity penetration, at around 5.2~s, the $D_\alpha$ and Mirnov signals show large, isolated bursts with an ELM repetition frequency of about 50~Hz, identifying them as Type-I ELMs; this classification is consistent with the ELITE linear stability result for the pre-seeding pedestal. At around 6.0~s, the burst amplitude is strongly reduced while the repetition frequency increases to about 600~Hz, indicating a transition to Type-III ELMs. By 6.27-6.30~s, both $D_\alpha$ spikes and Mirnov bursts are strongly suppressed, marking the ELM-free stage.

\begin{figure}[H]
    \centering
    \includegraphics[width=0.9\linewidth]{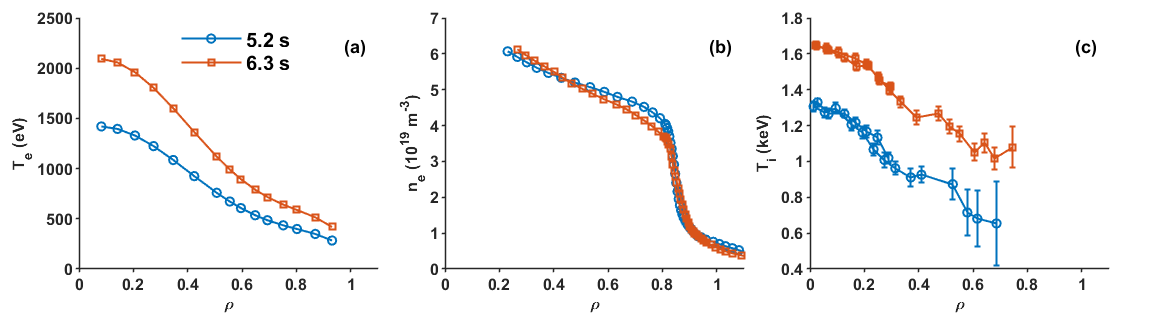}
    \caption{Comparison of electron temperature, electron density and ion temperature profiles for EAST discharge \#153901 at 5.2~s (blue) and 6.3~s (red). The profiles show a pronounced increase in the electron and ion temperatures, together with a modest relaxation of the pedestal density gradient, after nitrogen entered the plasma.}
    \label{fig:comparison}
\end{figure}

The profile comparison in Fig.~\ref{fig:comparison} shows that ELM suppression is accompanied by improved confinement rather than pedestal degradation. Both electron and ion temperatures increase after nitrogen entered the plasma, whereas the electron density pedestal is slightly relaxed. This channel-dependent profile evolution is consistent with the increase in stored energy and $H_{98(y2)}$ shown in Fig.~\ref{fig:discharge}. It also argues against a simple pressure-gradient-clamping interpretation of ELM suppression. The following sections therefore focus on the edge coherent mode that appears during this high-confinement ELM-free phase and on its role in the nonlinear ELM-suppression mechanism.

\subsection{Spectral features and radial localization}
\begin{figure}[H]
    \centering
    \includegraphics[width=0.7\linewidth]{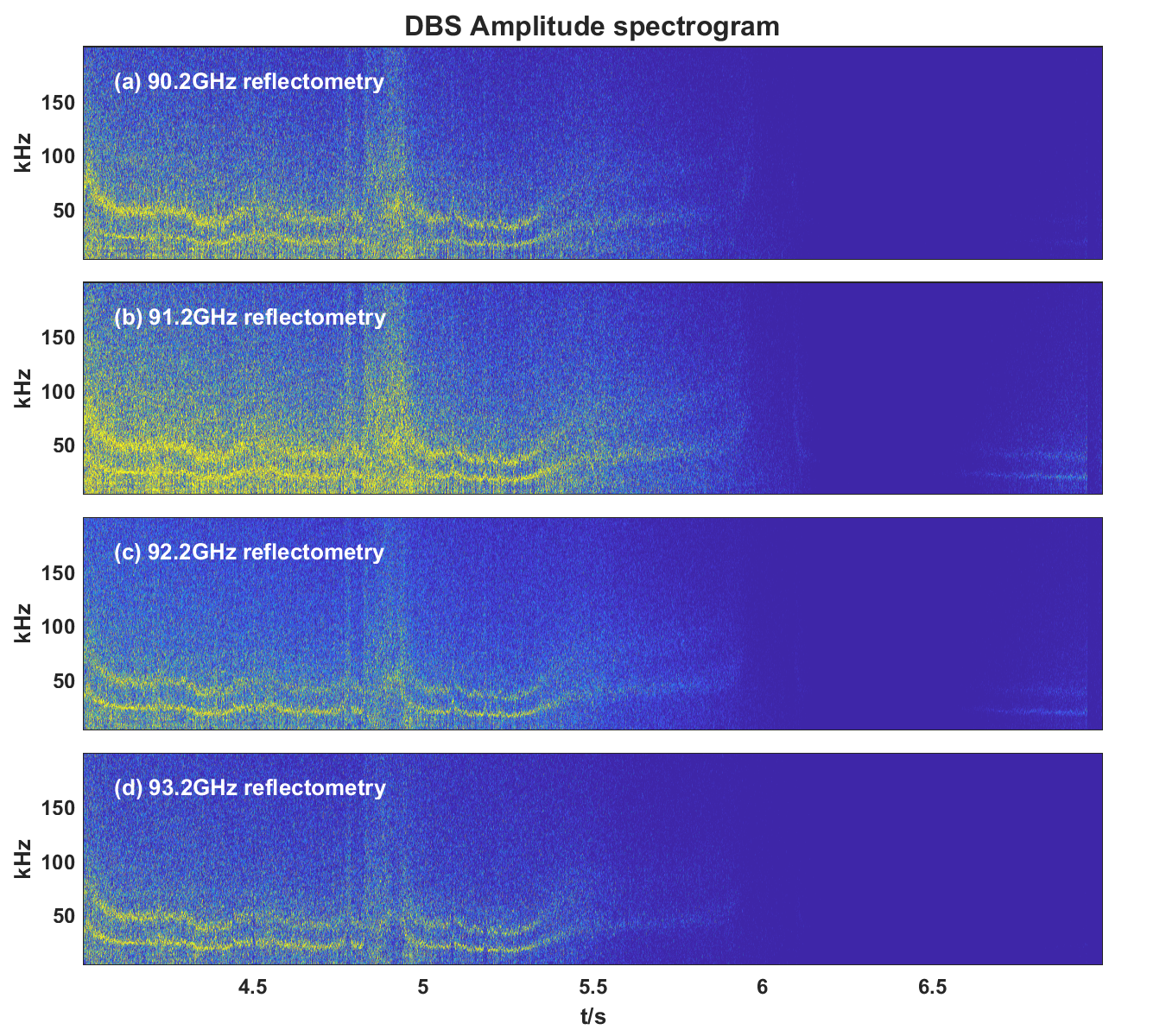}
    \caption{Doppler Backscattering}
    \label{DBS_AMP}
\end{figure}

The Doppler backscattering (DBS) system operated in X-mode at 90.2-94.2 GHz and sampled the pedestal top and steep-gradient region. Before impurity penetration, the DBS spectrogram in Fig.~\ref{DBS_AMP} shows a pre-existing edge coherent mode (ECM) at 20-90 kHz. This mode is localized near the steep-gradient pedestal region ($\psi_N \approx 0.95$) and is consistent with the electrostatic ECM commonly observed in EAST H-mode plasmas, where it is identified as Trapped Electron Mode (TEM)~\cite{guo2014recent, yu2025identification}.

\begin{figure}[H]
    \centering
    \includegraphics[width=0.75\linewidth]{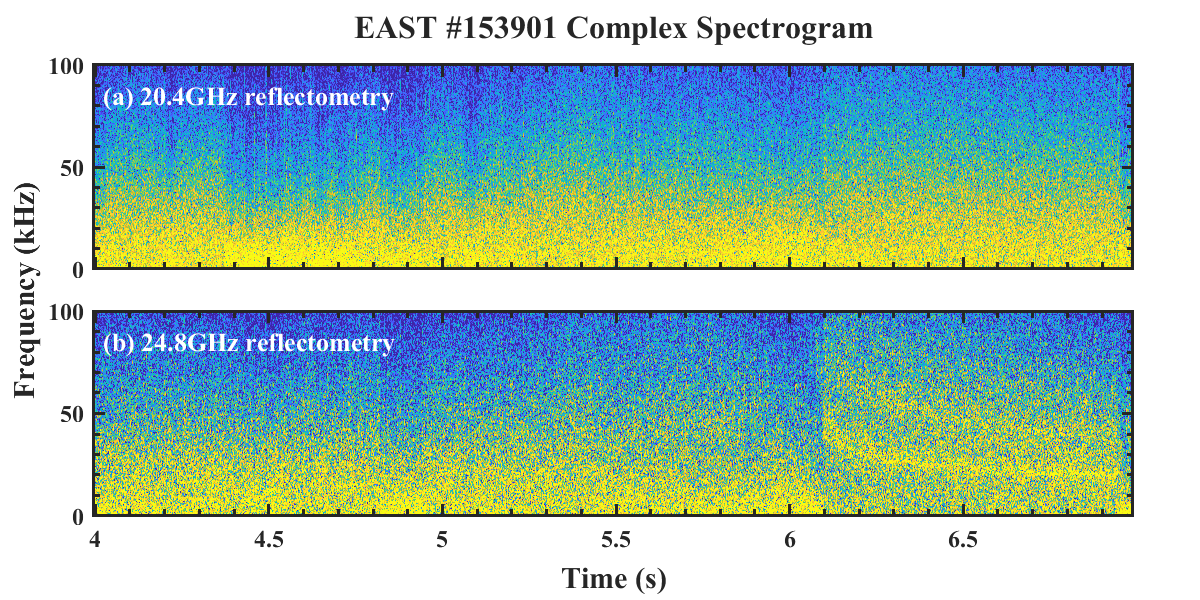}
        \caption{K/Ka band of the Poloidal Correlation Reflectometry, the detected complex signal is expressed as $I+iQ$. }
        \label{fig:Refl_Kka}
\end{figure}

After nitrogen entered the plasma, broadband pedestal turbulence in the DBS amplitude spectrogram was almost completely suppressed, and the pre-existing ECM disappeared. At the same time, the PCR signals in Fig.~\ref{fig:Refl_Kka} show the emergence of a new coherent fluctuation at the pedestal foot in the 24.8 GHz channel. The mode frequency chirps downward from approximately 40~kHz to 20~kHz, and its onset coincides with complete ELM suppression.

O-mode reflectometry provides an initial radial constraint on this new mode. The coherent fluctuation is clearly detected by the 24.8~GHz channel but is absent from the 20.4~GHz channel. Because each channel is most sensitive to density fluctuations near its cutoff layer, this contrast places the mode between the corresponding cutoff layers, namely near the pedestal foot.

\begin{figure}[H]
    \centering
    \includegraphics[width=0.7\linewidth]{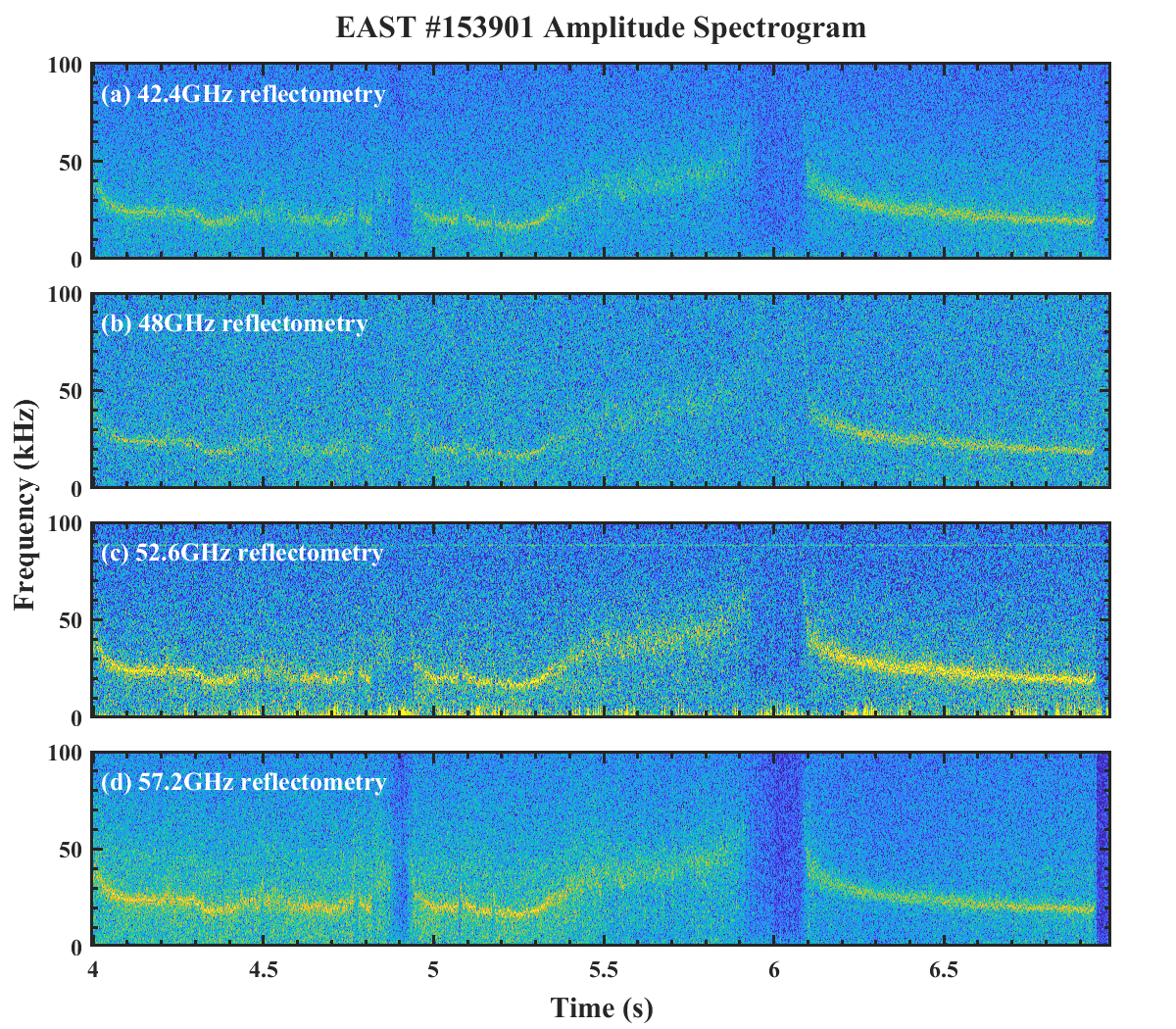}
        \caption{U band of the Poloidal Correlation Reflectometry}
        \label{fig:refl_U}
\end{figure}

Further inward, the U-band reflectometry data in Fig.~\ref{fig:refl_U} further separate the pre- and post-suppression modes. These inner channels detect both the original ECM before ELM suppression and the new coherent mode after suppression. The post-suppression mode has a larger fluctuation amplitude than the original ECM, but it is observed at a more outward cutoff region. Since the original ECM is detected by the inner probing channels but not by the 24.8~GHz channel, it is localized between the 42.4~GHz and 24.8~GHz cutoff layers, closer to the steep-gradient pedestal region than the new pedestal-foot mode. 

\begin{figure}[H]
    \centering
    \includegraphics[width=0.85\linewidth]{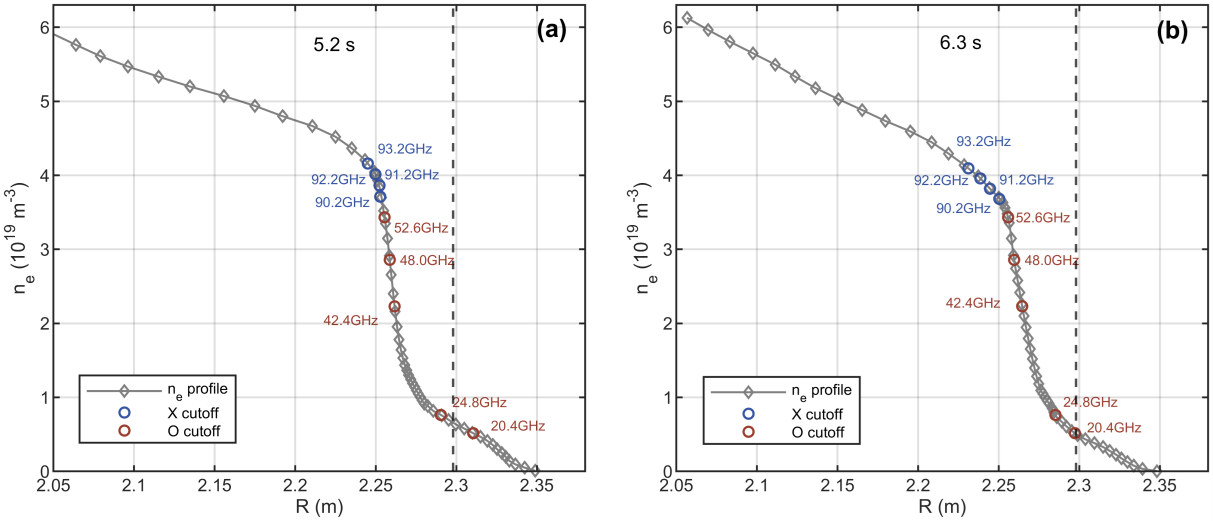}
    \caption{Density profile measured by reflectometry, with the locations of the cutoff corresponding to different frequencies.}
    \label{fig:density profiles}
\end{figure}

To convert these frequency constraints into radial locations, the reflectometry probing frequencies are mapped onto the measured electron-density profiles. As shown in Fig.~\ref{fig:density profiles} for t=5.2~s and t=6.3~s, the cutoff layers for the higher-frequency channels (33-52.6~GHz) lie in the steep-gradient region, whereas the 24.8~GHz cutoff is located near the pedestal foot ($R \approx 2.29~$m, $\psi_N \approx 0.98-0.99$). The 20.4~GHz cutoff is farther outward in the scrape-off layer and shows no coherent fluctuation. Taken together, the DBS, K/Ka-band and U-band reflectometry measurements localize the new 20-50~kHz mode to a narrow pedestal-foot layer.

\begin{figure}[H]
    \centering
    \includegraphics[width=0.75\linewidth]{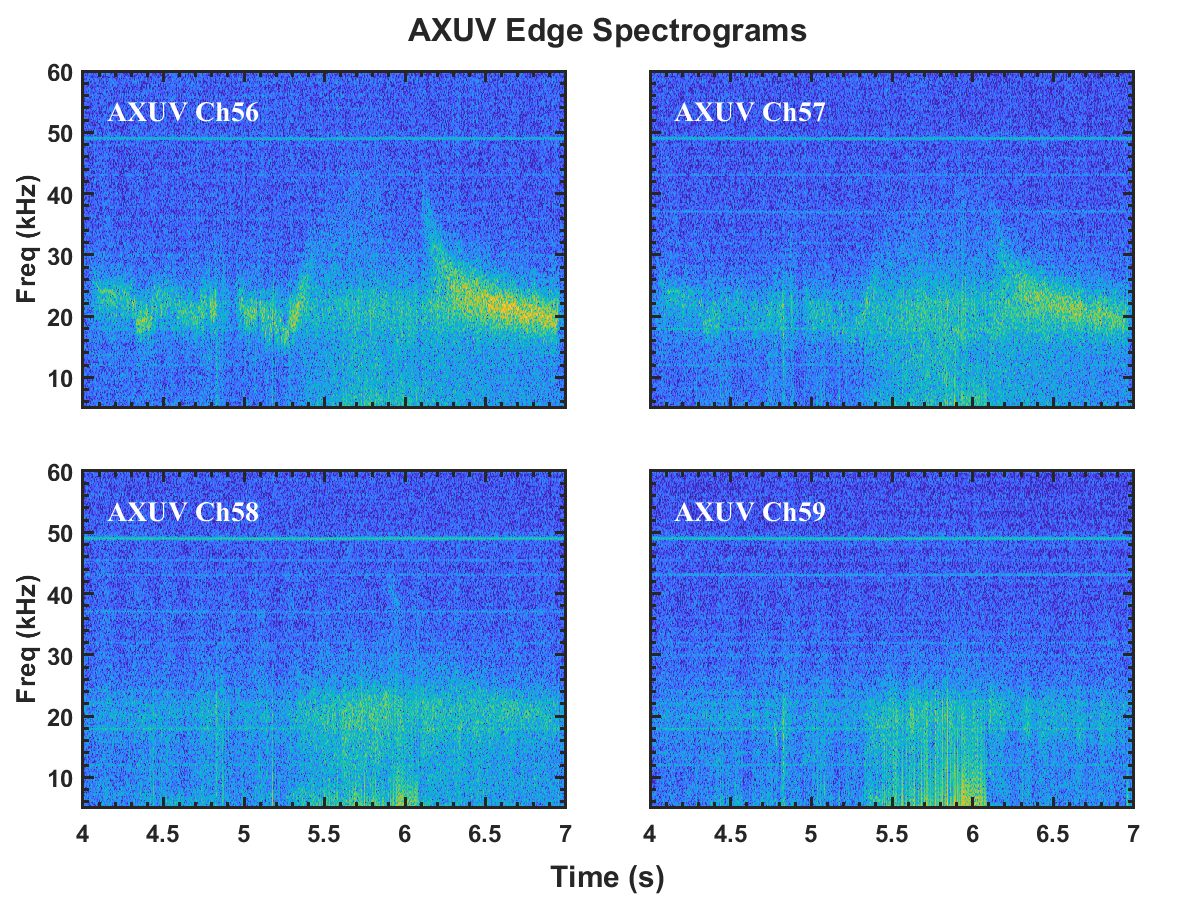}
    \caption{Spectrograms of edge AXUV}
    \label{fig:AXUV_edge}
\end{figure}

The AXUV bolometer arrays provide an independent check of this localization. Because the AXUV signals are line-integrated along the viewing chords, the radial position of a localized mode can be constrained by comparing adjacent channels. As shown in Fig.~\ref{fig:AXUV_edge}, the coherent fluctuation is visible in channels 56 and 57 but is absent from the outer channels 58 and 59. The mode is therefore constrained to lie between the tangent radii of AXUV channels 57 and 58, consistent with the pedestal-foot location inferred from reflectometry.

This cross-diagnostic consistency is important. Reflectometry excludes the steep-gradient region and the scrape-off layer as the primary location of the new mode, while the AXUV channel pattern independently confines it to the same edge layer. The nitrogen-induced coherent mode is therefore displaced outward from the pre-seeding ECM and resides in the pedestal foot.

The AXUV spectrograms in Fig.~\ref{fig:AXUV_edge} also show that the post-seeding coherent fluctuation has a larger radiative fluctuation amplitude than the pre-seeding ECM.

\begin{figure}[H]
    \centering
    \includegraphics[width=0.8\linewidth]{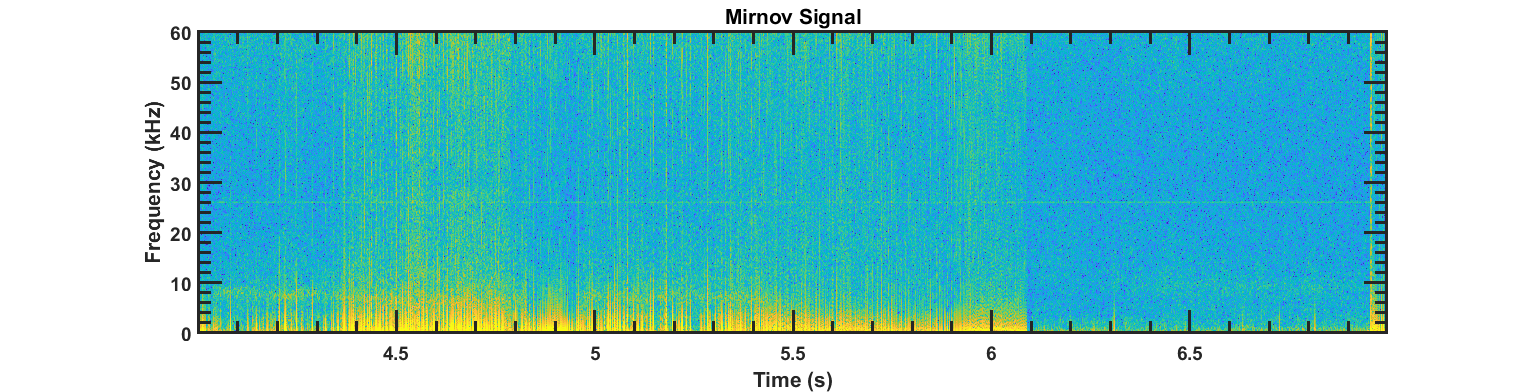}
    \caption{Mirnov Signal}
    \label{fig:mirnov spect}
\end{figure}

An analysis of the high-frequency Mirnov coil signals in Fig.~\ref{fig:mirnov spect} reveals no corresponding magnetic fluctuation in the 20-50 kHz range. This magnetic-null result indicates that the observed coherent mode is predominantly electrostatic.

We therefore conclude that nitrogen seeding does not simply strengthen the pre-existing steep-gradient ECM. Instead, it replaces it with a new, radially outward and predominantly electrostatic coherent mode at the pedestal foot ($\psi_N \approx 0.98-0.99$). This localization and magnetic-null result distinguish the observed mode from typical electromagnetic EHO/ECM activity located closer to the steep-gradient pedestal~\cite{burrell2001quiescent}.

\subsection{Mode number analysis and dispersion relation}

\begin{figure}[H]
    \centering
    \begin{subfigure}[h]{0.55\linewidth}
        \includegraphics[width=\linewidth, height=8cm, keepaspectratio]{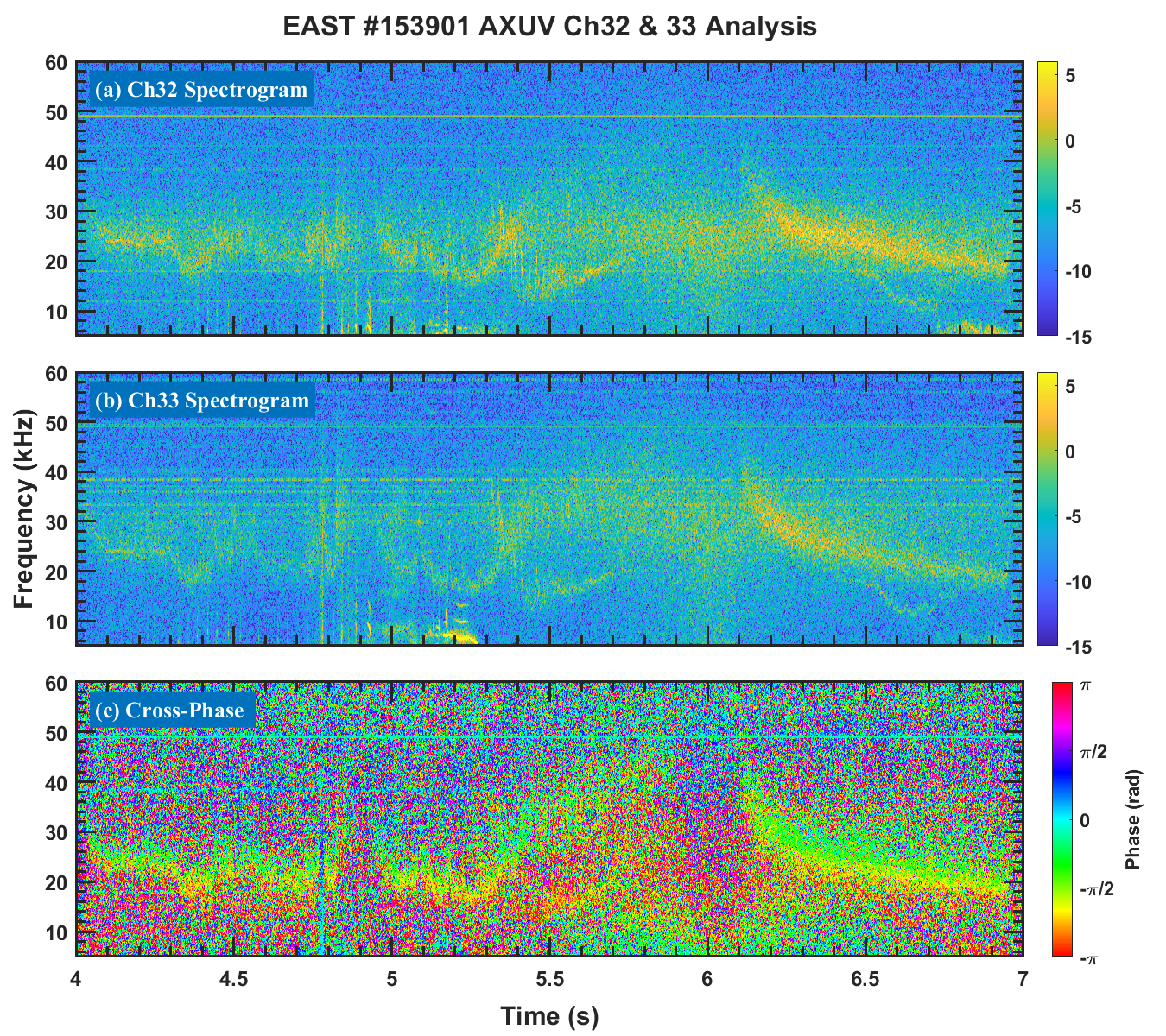}
        \caption{AXUV spectrogram}
        \label{fig:AXUV_spec}
    \end{subfigure}
    \hfill
    \begin{subfigure}[h]{0.4\linewidth}
        \includegraphics[width=\linewidth, height=8cm, keepaspectratio]{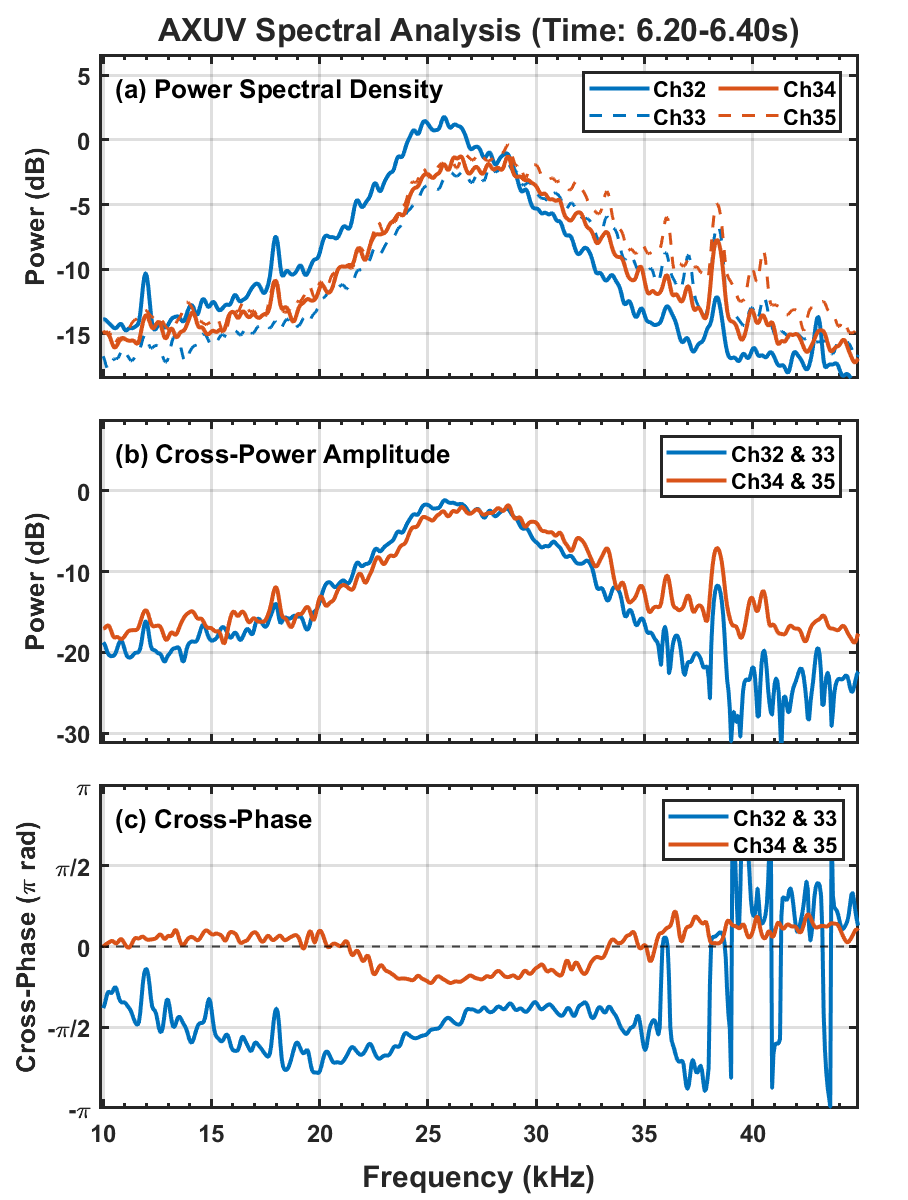}
        \caption{Cross Spectrum for 6.2s-6.4s}
        \label{fig:AXUV_cross}
    \end{subfigure}
    
    \caption{AXUV Correlation Analysis}
    \label{fig:AXUV}
\end{figure}

After the radial localization was established, the poloidal wavenumber $k_\theta$ and the mode numbers  $(m,n)$ were estimated from the cross-phase between adjacent AXUV midplane channels, as shown in Fig.~\ref{fig:AXUV_spec}. The calculation uses local quantities at the mode position: the safety factor $q \approx 9.3$ from magnetic equilibrium reconstruction, the electron temperature $T_e \approx$ 180~eV from the measured profiles (Fig.~\ref{fig:comparison}), and the local magnetic field $B \approx$ 5.3~T. The relevant separation $\Delta l_{\mathrm{pol}}$ is the distance between the two intersection points where the localized mode crosses the two AXUV viewing chords on the same flux surface, rather than a purely geometrical chord projection. The poloidal wavenumber is then derived as

\begin{equation}
k_\theta = \frac{\Delta \Phi}{\Delta l_{\mathrm{pol}}},
\end{equation}

where $\Delta \Phi$ is the measured phase shift at the center frequency of the mode. From the cross-power spectrum in Fig.~\ref{fig:AXUV_cross}(c), the phase shift at the dominant frequency is obtained as $\Delta \Phi \approx 0.3\pi$; the negative sign of the raw phase shift simply indicates the propagation direction of the mode.

Using these parameters, the poloidal wavenumber is $k_\theta \approx 0.538~$cm$^{-1}$. With the local magnetic pitch angle, this gives a toroidal mode number $n \approx 13$. Combining this with the local safety factor $q = m/n \approx 9.3$ gives a poloidal mode number $m \approx 123$. The wavenumber is then normalized to the ion sound Larmor radius, $\rho_s = \sqrt{m_i T_e} / eB$, which is approximately 0.0366~cm for the local plasma parameters. The resulting normalized wavenumber is $k_\theta \rho_s \approx 0.0197$.

These measurements establish the key experimental constraints on the mode: it is localized at the pedestal foot, predominantly electrostatic, and characterized by a low normalized wavenumber. However, these observables alone do not uniquely identify the underlying micro-instability, such as a trapped electron mode or an ion temperature gradient mode. The physical nature of the mode and the role of the local driving gradients are therefore examined using linear gyrokinetic simulations in the following section.

\section{Characterization of the pedestal foot mode with CGYRO}

\subsection{Plasma profiles and simulation setup}

To identify the micro-instabilitiy associated with the coherent mode at the pedestal foot, linear gyrokinetic simulations were performed with CGYRO. The calculations used experimentally constrained local profiles from EAST discharge \#153901 during the stationary ELM-free phase at $t = 6.3$~s. The simulation surface was chosen at $\psi_N = 0.99$, where the 20–50~kHz coherent fluctuation is localized by the reflectometry and AXUV measurements.

At this radial location, the local plasma parameters are characterized by steep electron temperature and density gradients, with the normalized electron temperature gradient $a/L_{T_e} = 151.87$ and the normalized electron density gradient $a/L_{n_e} = 44.13$, and by a local safety factor $q = 9.29$. Because the surface is close to the separatrix, the magnetic shear is very high ($\hat{s} \approx 65$). Nitrogen seeding also increases the effective charge to  $Z_{eff} \approx 2.03$, giving a relatively high normalized electron collisionality $\nu_e \approx 1.13$. These local conditions provide the input needed to test whether the measured ECM is compatible with a collisional trapped-electron instability.

The kinetic profiles used in the simulations were constrained by multiple diagnostics. The electron density and temperature profiles were obtained from reflectometry and Thomson scattering, with consistency checked against ECE measurements. The ion temperature profile was provided by X-ray crystal spectroscopy (XCS), and the magnetic equilibrium was reconstructed with EFIT and cross-validated by ONETWO integrated modeling. Both electrostatic and electromagnetic perturbations were retained in the calculation, and a high poloidal grid resolution along the extended ballooning angle was used to resolve the mode structure in this high-collisionality, high-shear edge region.

\subsection{Linear Growth Rate and Eigenfunction Analysis}
\begin{figure}[H]
    \centering
    \includegraphics[width=0.75\linewidth]{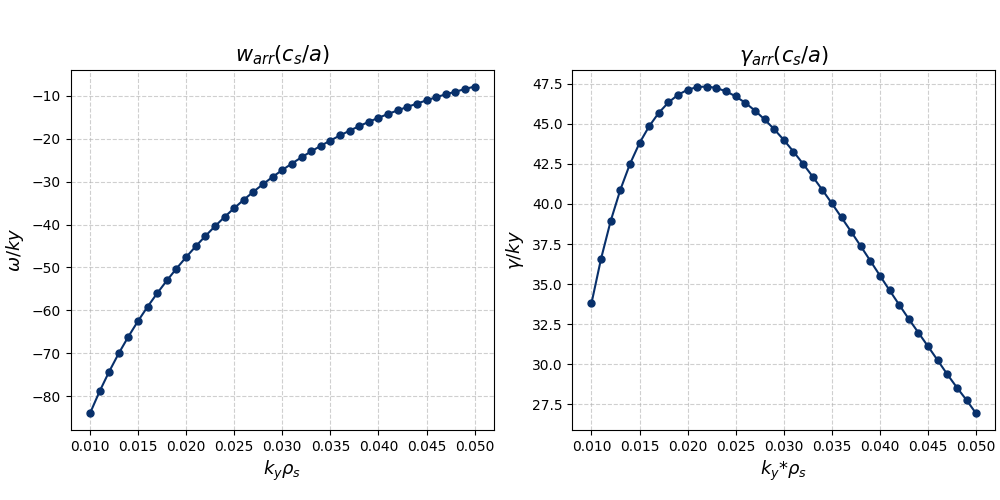}
    \caption{ Real frequency $\omega$ and normalized linear growth rate $\gamma$ spectra as functions of the normalized wavenumber $k_y \rho_s$ at $\psi_N = 0.99$ for $t = 6.3$\,s.}
    \label{fig:CGYRO_k_scan}
\end{figure}

The linear stability characteristics of the observed mode were then examined with CGYRO. Fig.~\ref{fig:CGYRO_k_scan} shows the spectra of the normalized linear growth rate $\gamma$ and real frequency $\omega$ as functions of the normalized binormal wavenumber $k_y \rho_s$. The real frequency is negative over the relevant wavenumber range, indicating propagation in the electron diamagnetic drift direction. This propagation direction is inconsistent with an ion temperature gradient (ITG) mode~\cite{kumar2025itg,jian2025shafranov}.

The dominant instability peaks at a low wavenumber, $k_y \rho_s \approx 0.022$, corresponding to a physical wavenumber $k_y \approx 0.47\,\mathrm{cm}^{-1}$. This value agrees well with the experimentally inferred poloidal wavenumber $k_\theta \approx 0.54\,\mathrm{cm}^{-1}$ from AXUV cross-phase analysis. The low$-k_y$ scale is also far below the range expected for electron temperature gradient (ETG) modes, which typically occur at $k_y\rho_s$ of order unity or higher~\cite{howard2024simultaneous,kumar2025itg,li2025role}.

\begin{figure}[H]
    \centering
    \includegraphics[width=0.75\linewidth]{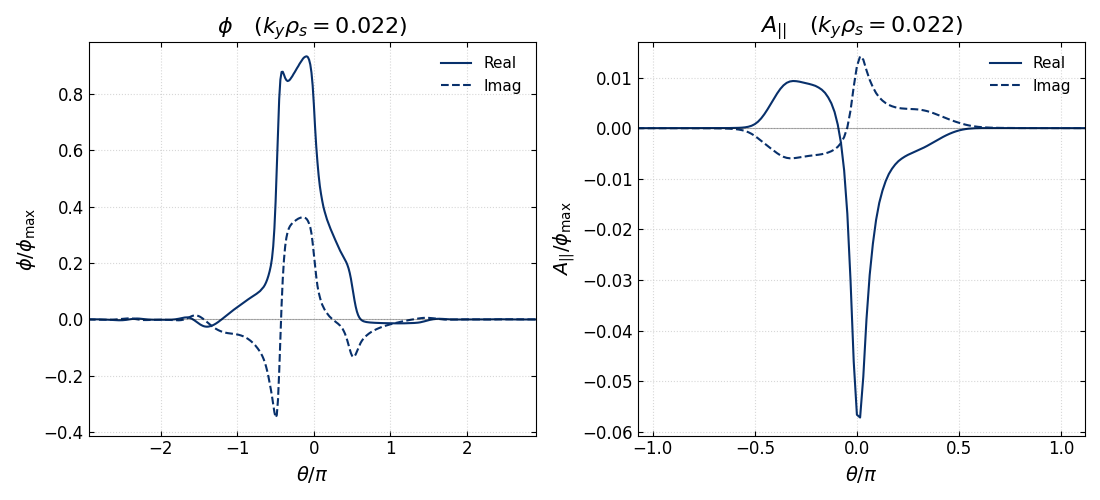}
    \caption{Eigen Functions of k=0.022}
    \label{fig:eigenfunctions}
\end{figure}

The eigenfunctions of the electrostatic potential $\phi$ and the parallel magnetic vector potential $A_\parallel$ provide an additional constraint on the mode identification ((Fig.~\ref{fig:eigenfunctions})~\cite{howard2024simultaneous}. The potential $\phi$ has a localized, even-parity structure centered near the outboard midplane ($\theta = 0$), while $A_\parallel$ is much weaker and shows a mixed-parity structure. Together with the electron-directed propagation and the low normalized wavenumber, these eigenfunction properties identify the instability as a trapped electron mode (TEM). Furthermore, the dominantly electrostatic character of the calculated mode is also consistent with the absence of a corresponding 20–50~kHz Mirnov signal in the experiment.

\subsection{Parametric Dependence and Identification of DTEM}

\begin{figure}[H]
    \centering
    \begin{subfigure}[b]{0.32\textwidth}
        \centering
        \includegraphics[width=\textwidth]{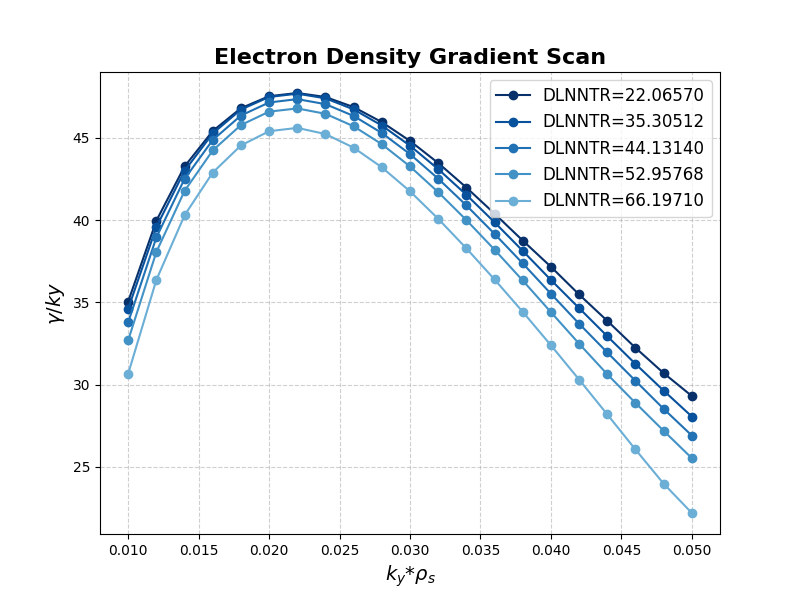}
        \caption{Density gradient scan}
        \label{fig:6.3s_N_scan}
    \end{subfigure}
    \hfill 
    \begin{subfigure}[b]{0.32\textwidth}
        \centering
        \includegraphics[width=\textwidth]{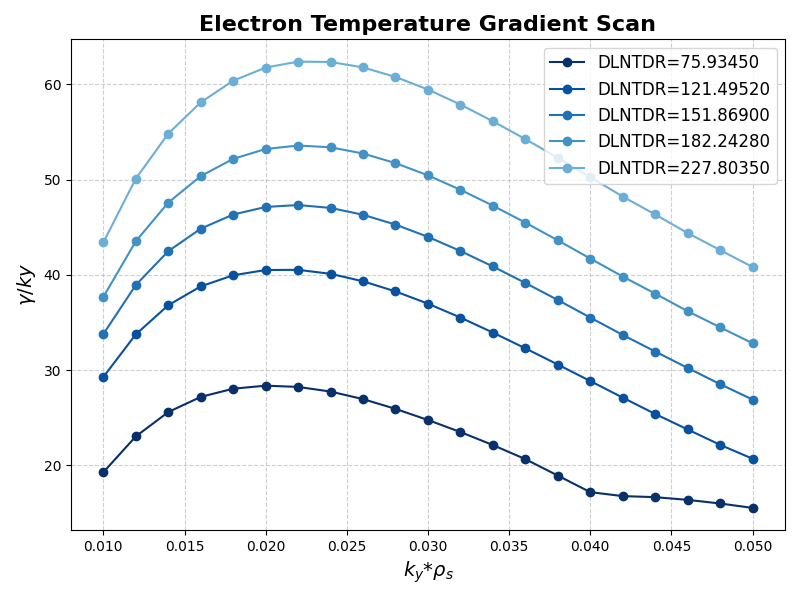}
        \caption{Temperature gradient scan }
        \label{fig:6.3s_T_scan}
    \end{subfigure}
    \hfill
    \begin{subfigure}[b]{0.32\textwidth}
        \centering
        \includegraphics[width=\textwidth]{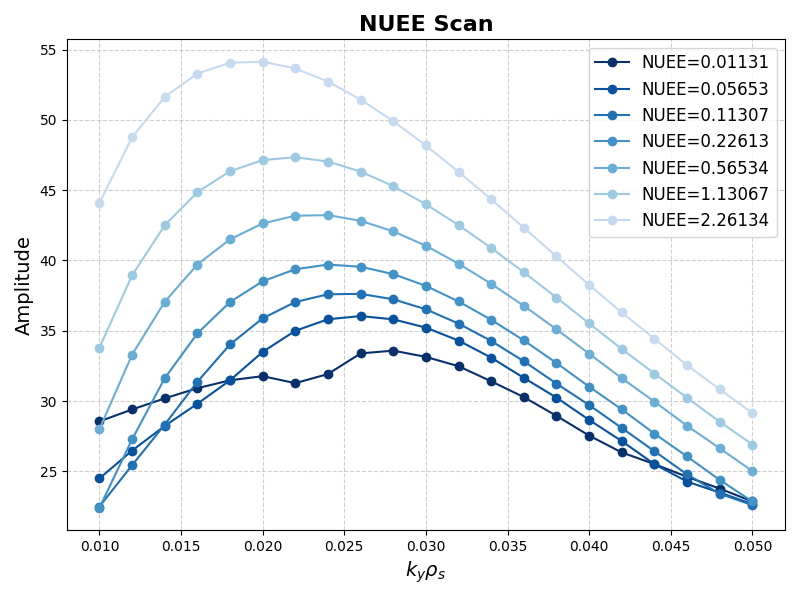}
        \caption{Collisionality scan }
        \label{fig:6.3s_NUEE_scan}
    \end{subfigure}
    \caption{Parametric dependencies of the normalized linear growth rate at $k_y \rho_s = 0.022$ on (a) normalized density gradient $a/L_{n_e}$, (b) normalized temperature gradient $a/L_{T_e}$, and (c) normalized electron collisionality $\nu_e$.}
    \label{fig:cgyro_scan}
\end{figure}

To further classify the instability, parametric scans were performed in the normalized density gradient $a/L_{n_e}$, electron temperature gradient $a/L_{T_e}$ and electron collisionality $\nu_e$. As shown in Fig.~\ref{fig:6.3s_N_scan} and~\ref{fig:6.3s_T_scan}, the linear growth rate significantly increases with electron temperature gradient, but is weakly related to density gradient. This indicates that the mode is a TEM driven by electron temperature gradient~\cite{howard2021gyrokinetic}.

The collisionality scan in Fig.~\ref{fig:6.3s_NUEE_scan} provides the key distinction between a collisionless TEM and a dissipative TEM. Increasing $\nu_e$ further destabilizes the mode, whereas collisions would normally reduce the drive of a collisionless TEM by detrapping electrons. This positive collisionality dependence, together with the electrostatic eigenfunction and the absence of a magnetic Mirnov counterpart, supports identification of the observed pedestal-foot ECM as a dissipative trapped electron mode (DTEM), rather than a micro-tearing mode~\cite{howard2021gyrokinetic,jian2019role}.

Overall, the CGYRO results reproduce the main experimental constraints on the ECM: the low normalized wavenumber, electron-diamagnetic propagation, predominantly electrostatic character and sensitivity to the nitrogen-modified edge collisionality and gradients. The simulations therefore identify the 20–50~kHz pedestal-foot ECM as a DTEM-like instability excited under the local edge conditions after nitrogen seeding.

\section{Discussion and Conclusion}

The combined measurements show that nitrogen seeding substantially changes the edge fluctuation state while maintaining improved confinement. After nitrogen enters the plasma, the pre-existing ECM in the steep-gradient region disappears and a new 20–50~kHz ECM appears at the pedestal foot. At the same time, the ELM activity evolves from large Type-I events to small, high-frequency Type-III ELMs and finally to an ELM-free state, while the temperature pedestal, stored energy and $H_{98(y,2)}$ increase.

A critical constraint comes from the ELITE linear stability analysis. In the ELM-free phase, the linear growth rate of the P-B instability is higher than that before ELM suppression. Therefore, the disappearance of large ELM crashes cannot be explained by a simple reduction of the pedestal pressure gradient or by moving the pedestal farther away from the P-B stability boundary. Instead, the observations indicate that the nitrogen-induced ECM modifies the nonlinear evolution of the P-B instability.  

This interpretation is consistent with coherent-mode-mediated ELM mitigation mechanisms reported in recent EAST simulations and multiscale edge studies~\cite{yu2025identification,li2025multi}, where an edge coherent mode can alter the growth, phase coherence, saturation or energy-transfer pathway of P-B modes through nonlinear interaction. 

Reflectometry and AXUV measurements identify the new 20–50~kHz ECM localized at the pedestal foot ($\psi_N \approx 0.98-0.99$), with no clear magnetic counterpart in Mirnov coils and with $k_\theta \approx$ 0.5-0.6~cm$^{-1}$. Linear gyrokinetic simulations with CGYRO reproduce the measured electron-direction propagation and low normalized wavenumber, $k_y \rho_s = 0.022$ (corresponding to $k_y \approx 0.47\,\mathrm{cm}^{-1}$), and parameter scans identify the mode as a dissipative trapped electron mode driven by the nitrogen-modified edge conditions.

Together with ELITE results showing an increased P-B linear growth rate in the ELM-free phase, these observations indicate that ELM suppression is not caused by simple weakening of the pedestal drive. Instead, the results support a coherent-mode-mediated nonlinear interaction between the pedestal-foot ECM and P-B dynamics. Future nonlinear BOUT++ simulations based on the measured pedestal profiles will therefore be needed to determine how this ECM suppresses large ELM crashes while preserving the improved-confinement pedestal.

\section*{Acknowledgments}
{This work is supported by the National MCF Energy R\&D Program under Grant No. 2024YFE\-03060004. We thank the EAST team (https://cstr.cn/31130.02.EAST ), for providing technical support and assistance in data collection and analysis. We also express our sincere gratitude to Li Yanlong for the valuable discussions regarding the BOUT++ simulations.}

\printbibliography

\end{document}